\documentclass[journal]{IEEEtran}
\usepackage{color}
\usepackage{amsmath}

\usepackage{graphics,graphicx}

\usepackage[numbers,sort&compress]{natbib}

\usepackage{bm}
\ifCLASSINFOpdf
\else
\fi

\begin{document}

\title{Quasi-Continuous Metasurfaces for Orbital Angular Momentum Generation}
%
%
%

\author{Menglin~L.~N.~Chen,~\IEEEmembership{Member,~IEEE,}
        Li~Jun~Jiang,~\IEEEmembership{Fellow,~IEEE,}
        and~Wei~E.~I.~Sha,~\IEEEmembership{Senior~Member,~IEEE}
\thanks{M. L. N.~Chen and L. J. Jiang are with the Department of Electrical and Electronic Engineering, The University of Hong Kong, Hong Kong (e-mail: menglin@connect.hku.hk; jianglj@hku.hk).}
\thanks{W.~E. I.~Sha is with the Key Laboratory of Micro-nano Electronic Devices and Smart Systems of Zhejiang Province, College of Information Science and Electronic Engineering, Zhejiang University, Hangzhou 310027, China (email: weisha@zju.edu.cn).}}

%
%

\markboth{Ieee antennas and wireless propagation letters,~Vol.~XX, 2019}%
{Shell \MakeLowercase{\textit{et al.}}: Bare Demo of IEEEtran.cls for IEEE Journals}
%



\maketitle

\begin{abstract}
A quasi-continuous composite perfect electric conductor-perfect magnetic conductor metasurface and a systematic metasurface design process are proposed for the orbital angular momentum (OAM) generation. The metasurfaces reflect the incident left circularly polarized (LCP)/right circularly polarized (RCP) plane wave to RCP/LCP vortex beams carrying OAM at normal or oblique direction. Unlike conventional metasurfaces that are composed of discrete scatterers, the scatterers on the proposed metasurface form a quasi-continuous pattern. The patterning of the metasurface is calculated through grating vectors, and no optimization of single scatterer is required. Furthermore, the distortions from local-response discontinuity of discrete scatterers are avoided. This letter provides great convenience to high-quality OAM generation.
\end{abstract}


\section{Introduction}

\IEEEPARstart{I}{t} is known since 1992 that Laguerre-Gaussian modes carry well-defined orbital angular momentum (OAM). Those modes have an azimuthally dependent phase factor, $e^{il\phi}$, where $\phi$ is the azimuthal angle, and $l$ is the OAM index~\cite{allen1992OAM}. Since then, OAM has been applied in communications~\cite{willner2015optical}, imaging~\cite{tamburini2006overcoming,liukang2015OAM}, and so on. For example, OAM-based multiplexing and demultiplexing have been implemented at different frequency regimes~\cite{thide2012encoding,yanyan2014high}. As a simple and newly discovered multiplexing approach, OAM multiplexing needs to be studied to overcome the issues of power decay, sensitivity to misalignment, and mode crosstalk for practical applications~\cite{yanyan2015performance,kahn2015capacity,2018vortex}. Furthermore, OAM is promising for applications related to the light-matter interaction, such as the manipulation of particles and detection of spinning objects~\cite{grier2003a,padgett2013detection}.

To introduce OAM to a plane wave, the azimuthal phase factor, $e^{il\phi}$ needs to be added along the wave path. Devices for OAM generation include spiral phase plate~\cite{woerdman1994helical}, q plates~\cite{paparo2006optical}, gratings~\cite{qiu2018spiniform}, metasurfaces~\cite{cuitiejun2018generation,lilong2016generating,shihongyu2018transparent} and photonic crystals~\cite{menglin_pc}. With high flexibility, metasurfaces have been widely used in wave manipulation~\cite{cuitiejun2014coding,menglin_AWPL,shihongyu2019generation}. Particularly, as one type of metasurfaces, geometric-phase based metasurface changes the local phase of electromagnetic (EM) waves by scatterers with varying orientations. This phase modification originates from the change in the polarization state along different paths on the Poincar\'{e} sphere~\cite{berry1987the}. The key point to produce OAM waves by using geometric-phase based metasurface is to make the local phase change equal to $e^{il\phi}$. Various prototypes of scatterers on these metasurfaces have been proposed for OAM generation, such as split-ring resonators~\cite{menglin_csrr,menglin2018oam}, L-shaped and elliptical nano antennas~\cite{boyd2014generating,luoxiangang2015a}. However, the performance could be degraded due to the unwanted coupling among scatterers and high-order diffraction. Hence, a lot of efforts need to be taken in the design and optimization of scatterers.

In this work, we present a quasi-continuous metasurface for OAM generation. The metasurface contains an anisotropic perfect electric conductor (PEC) layer and an isotropic perfect magnetic conductor (PMC) layer. It reflects the incident left circularly polarized (LCP)/right circularly polarized (RCP) plane wave to a RCP/LCP wave, along with a locally modulated geometric phase. The phase distribution on the whole metasurface satisfies $e^{i(l\phi+k_x x + k_y y)}$ so that an OAM of order $l$ can be generated at the k-space position $(k_x,k_y)$,  where $k_x$ and $k_y$ are the transverse wavenumbers. The PEC layer presents a quasi-continuous pattern and the PMC layer can be realized using any artificial high impedance surface.

\section{Methodology}
\begin{figure}[!t]
\centering
\includegraphics[width=\columnwidth]{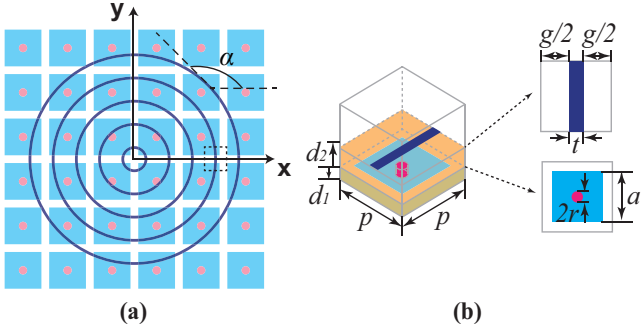}
\caption{Schematic pattern of the PEC-PMC metasurface. (a) Top view of the whole metasurface. The inclination angle of the metal strips is denoted by $\alpha$. (b) A scatterer in the metasurface.}
\label{grating_vector}
\end{figure}

Figure~\ref{grating_vector}(a) illustrates one type of composite PEC-PMC metasurfaces to produce vortex wave with $l=\pm2$~\cite{menglin_pecpmc}. The top PEC layer is composed of concentric metal loops. A mushroom-like high impedance surface acting as the PMC layer is put beneath the PEC layer. Any local area on the metasurface can be considered as a PEC scatterer on top of a PMC scatterer. One unit cell is extracted and shown in Fig.~\ref{grating_vector}(b). The geometric and material parameters are the same as in our previous work~\cite{menglin_pecpmc}. The size of one mushroom structure is around $\lambda_0/7$. Therefore, the periodically distributed mushroom can be considered as an isotropic, homogeneous plane. Complex Jones matrix ($J_{xx} ~ J_{xy}$; $J_{yx} ~ J_{yy}$) is used to model the scatterer. Since it is symmetric about the $x$ and $y$ axes, we have $J_{xy}=J_{yx}=0$. The metal-strip array behaves like a parallel-plate waveguide. The cut-off frequency of corresponding TE$_1$ mode is $1/(2g\sqrt{\mu \epsilon})$, where $g$ is the gap between adjacent metal strips~\cite{pozar}. Below the cut-off frequency, the $y$-polarized component will be totally reflected with a $\pi$ phase shift ($J_{yy}=-1$). The penetrated $x$-polarized component will be reflected by the PMC plane without phase shift ($J_{xx}=1$). When the composite structure is illuminated by a circularly polarized wave, the polarization state will be changed from left to right and vice versa. It should be noticed that the whole metasurface is composed of those scatterers with different orientations. Therefore, besides the polarization flip, the reflected wave is accompanied by an additional phase which is equal to $e^{\pm 2 i \alpha}$, where $\alpha$ is the inclination angle between the metal-strip tangent and $x$ axis. The plus/minus sign is taken when the incident wave is LCP/RCP. As can be read from the figure, $\alpha=\phi+\pi/2$. Thus, the reflected wave carries an OAM of order $\pm 2$. Usually, for geometric-phase based metasurfaces, rotation of the discrete scatterers breaks the periodicity along the $x$ and $y$ directions, resulting in the undesired mutual coupling between the nearest-neighbor scatterers. However, in the composite PEC-PMC metasurface, individual control of the two orthogonal polarizations is realized by the PEC and PMC layers, respectively. No rotation operation is required for the scatterers on the PMC layer. Meanwhile, induced current is continuously guided along the smoothly connected metal strips on the PEC layer. From this point of view, compared to the discrete geometric-phase based scatterers, the unwanted mutual coupling among the PEC-PMC scatterers is significantly reduced. Moreover, by keeping the local period small enough, no high-order diffraction exists.

\begin{figure}[!t]
\centering
\includegraphics[width=\columnwidth]{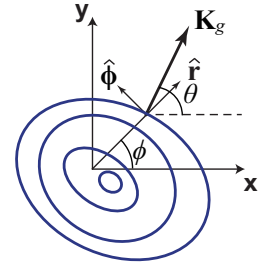}
	\caption{Arbitrary metal-strip pattern modeled by the grating vector $\mathbf{K}_g$ which is normal to the tangent of the metal strips.}
\label{grating_vector2}
\end{figure}

Regarding vortex beams carrying arbitrary orders of OAM, in conventional metasurfaces, discrete scatterers are placed at specific locations with pre-designed orientations~\cite{menglin_csrr,menglin2018oam,boyd2014generating,luoxiangang2015a}. However, the discontinuity and aperiodicity induced by the discrete scatterers will distort the near-field pattern~\cite{menglin_pecpmc} and lower the efficiency. To achieve a continuous phase shift on the metasurface, the discrete scatterers should be continuously and smoothly connected. For practical implementation, we consider the metal strips as a grating and model them using grating vectors. An arbitrary metal-strip pattern is drawn in Fig.~\ref{grating_vector2}. The grating vector is perpendicular to the tangent of metal strips. In other words, the grating vector is the normal vector of the curve of the metal strips. In polar coordinates, it is written as

\begin{align}
\begin{split}
	\mathbf{K}_{g}(r,\phi) &=  K_r\hat{\bm{r}}+  K_\phi \hat{\bm{\phi}} \\
	&=K_0(r,\phi) \cos[\theta(r,\phi)-\phi]\hat{\bm{r}} \\
	&+ K_0(r,\phi) \sin[\theta(r,\phi)-\phi]\hat{\bm{\phi}},
\end{split}
\label{eq:kg}
\end{align}
where $K_0(r,\phi)$ is the local spatial frequency at $(r,\phi)$.

As discussed, the orientation of the local metal segments characterized by $\theta$ determines the geometric phase, which will contribute to the phase shift for the wave manipulation. To produce vortex beams radiating at the desired direction, $\theta$ should satisfy
\begin{equation}
	\theta(r,\phi)=m \phi + a r \cos \phi  + b r \sin \phi  + D,
	\label{eq:theta}
\end{equation}
where $m=l/2$ and is also called the topological charge of the structure. The second and third terms in $\theta(r,\phi)$ are introduced for realizing oblique reflection. $a=k_x /2$, $b=k_y /2$. $D$ is a phase constant that determines the direction of $\mathbf{K}_g$ at the origin.

To ensure the continuity of the grating, it is required that $\nabla \times \mathbf{K}_g=0$~\cite{hasman2003formation}. By substituting~\eqref{eq:theta} into~\eqref{eq:kg}, the general solution of $K_0$ fulfilling the zero divergence condition is of the form
\begin{equation}
	K_0=A \frac{e^{a r \sin \phi  - b r \cos \phi }}{r^m},
	\label{eq:k0}
\end{equation}
where $A$ is a scaling factor.

It is supposed that the closed metal strip could be mathematically described by a curve equation $g(r,\phi)=C$.  $\nabla g$ is the normal vector of the curve. Therefore, $\nabla g = \mathbf{K}_{g}$ and it is solved by integrating $\mathbf{K}_{g} $ over an arbitrary path:
\begin{equation}
	g(r,\phi)= \int_{(r_0,\phi_0)}^{(r,\phi_0)} K_r dr+ \int_{(r,\phi_0)}^{(r,\phi)} r K_\phi d\phi.
	\label{eq:g}
\end{equation}

\begin{figure}[!t]
\centering
\includegraphics[width=\columnwidth]{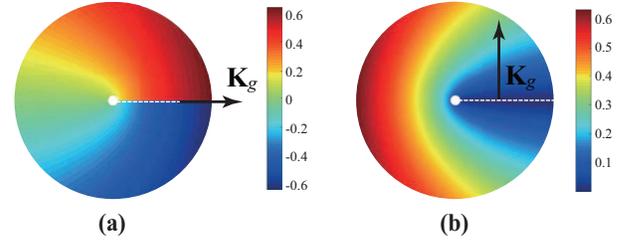}
\caption{The calculated $g$ when (a) $m=0.5$, $a=b=0$, $D=0$, $A=1$ and (b) $m=0.5$, $a=b=0$, $D=\pi/2$, $A=1$.}
\label{phi_m_0.5}
\end{figure}

In Fig.~\ref{phi_m_0.5}, we show the calculated $g$ when $\theta(r,\phi)=m \phi+D$. As expected, the orientation of the contour line coincides with $\theta(r,\phi)$. The grating vector is perpendicular to the contour line and its initial orientation at $\phi=0$ is determined by $D$ which is horizontal when $D=0$ and vertical when $D=\pi/2$.

\begin{figure}[!t]
\centering
\includegraphics[width=\columnwidth]{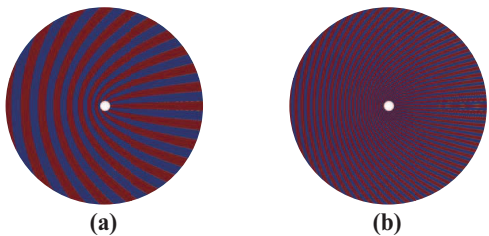}
\caption{The binary grating generated from the calculated $g$ in Fig.~\ref{phi_m_0.5}(b) when the scaling factor (a) $A=100$ and (b) $A=300$.}
\label{phi_m_0.5_s}
\end{figure}

A Lee-type binary grating is then generated from $g$~\cite{Lee1974binary}. Figure~\ref{phi_m_0.5_s} shows two gratings with different scaling factors $A$. The grating is a desired pattern for the PEC layer. The local orientations of the two patterns are identical while the local periods are different. Since adjacent metal strips function as a parallel-plate waveguide, proper value of $A$ needs to be chosen so that their gap is small enough to guarantee that the operating frequency is far below the cut-off frequency of the TE${_1}$ mode.

\section{Simulation}
\begin{figure}[!t]
\centering
\includegraphics[width=\columnwidth]{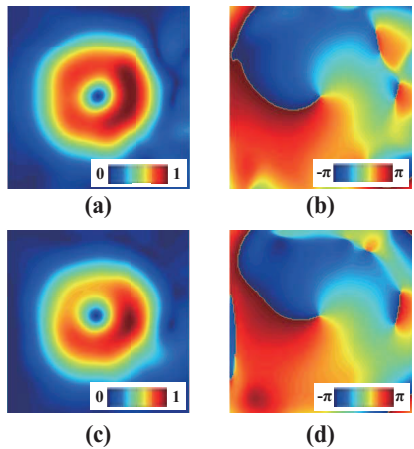}
\caption{Simulated field distributions at a transverse plane $20$~mm away from the metasurface with $m=0.5$.  (a) Amplitude and (b) phase of the field when the metal strips are placed on an ideal PMC boundary; (c) amplitude and (d) phase of the field when the metal strips are placed on the mushroom-like high impedance surface.}
\label{q_0.5}
\end{figure}

For the PEC layer in Fig.~\ref{phi_m_0.5_s}(b), the outer radius of the pattern is $100$~mm and the inner radius is $5$~mm. Two scenarios are compared in which the ideal PMC boundary and the mushroom-like high impedance surface are applied beneath the PEC layer, respectively. Simulations are done in CST MWS.  The metasurface is illuminated by a LCP Gaussian wave with the beam waist of $50$~mm at $6.2$~GHz. Figure~\ref{q_0.5} shows the amplitude and phase distributions of the reflected RCP field component at a transverse plane $20$~mm away from the metasurface. We see the field distribution of an OAM wave with $l=1$. A clear phase singularity is observed. The phase encounters a total $2\pi$ change along a closed path enclosing the center. The simulated fields in the two scenarios show a good agreement with each other.

\begin{figure}[!t]
\centering
\includegraphics[width=\columnwidth]{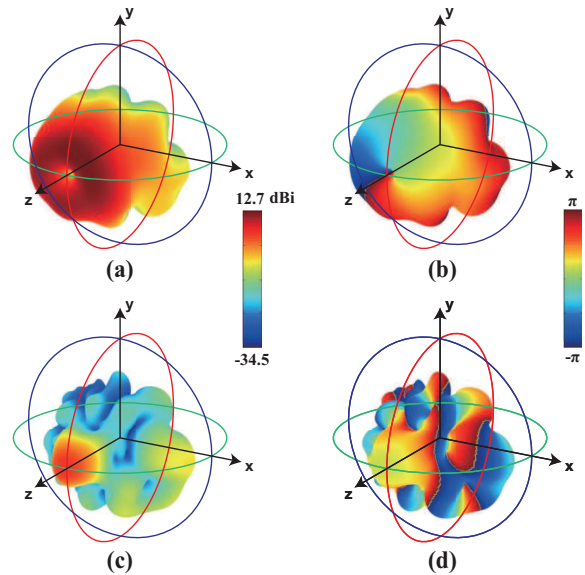}
\caption{Simulated far-field radiation patterns for the metasurface with $m=0.5$. (a) Directivity and (b) phase of the cross-polarized (RCP) component; (c) directivity and (d) phase of the co-polarized (LCP) component.}
\label{q_0.5_ff}
\label{cst_field}
\end{figure}

The efficiency of the unit cell can be used to evaluate the performance of the design. The parameters for the mushroom-like high impedance surface keep same for all metasurfaces. Hence, we can assume $J_{xx}=1$ all the time. The $y$-polarized component will be totally reflected due to presence of the ground plane. However, the reflection phase depends on the gap between the metal strips. Here, we write $J_{yy}=e^{ip}$. For the incidence of a circularly polarized wave, incident components are $i_{x}=1$, $i_{y}=\pm i$. The reflected wave components $r_{x}=J_{xx}=1$, $r_{y}=\pm i J_{yy}= \pm i e^{ip}$. The amplitude of the converted cross-circularly polarized component is $a=(r_x \pm i r_y)/\sqrt{2}= (1 - e^{ip}) / \sqrt{2}$. The efficiency of the unit cell, which is equal to the ratio of the power of the converted wave to that of the incident wave, is $|1-e^{ip}|^2/4$. Clearly, for the ideal case that $p=\pi$, the efficiency is $1$. The metasurface shown in Fig.~\ref{grating_vector} has been demonstrated to exhibit nearly perfect efficiency~\cite{menglin_pecpmc}. For the metasurface with non-uniformly distributed metal strips, we calculate the efficiency using the power of the reflected RCP wave divided by that of the incident wave. The efficiency of the metasurface for Fig.~\ref{q_0.5} is $94\%$. The far-field radiation patterns of both the co-polarized component and cross-polarized component are plotted in Fig.~\ref{q_0.5_ff}. The maximum directivity for the RCP component is $12.7$~dBi and it is only $5.5$~dBi for the LCP component. Moreover, the directivity for the converted component presents a donut shape with a gradual phase change around the vortex axis while the co-polarized component does not show such feature.

\begin{figure}[!t]
\centering
\includegraphics[width=\columnwidth]{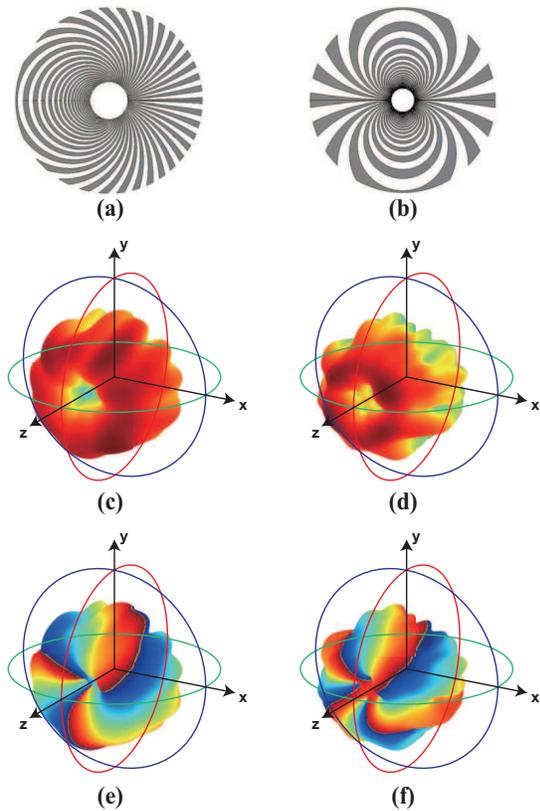}
\caption{The quasi-continuous PEC-PMC metasurfaces to produce OAM at normal direction and their far-field radiation patterns. When $m=1.5$, (a) the top view of the PEC layer; (c) directivity and (e) phase of the RCP component. When $m=2$, (b) the top view of the PEC layer; (d) directivity and (f) phase of the RCP component. }
\label{q_oam}
\end{figure}

In following simulations, we will use the ideal PMC boundary for simplicity. Metasurfaces for generating OAM with other orders can be built based on the same procedure. The top metal-strip layers with the topological charge $m=1.5$ and $m=2$ are shown in Fig.~\ref{q_oam}. The far-field directivity and phase patterns indicate the successful generation of OAM with order $l=3$ and $l=4$.

\begin{figure}[!t]
\centering
\includegraphics[width=\columnwidth]{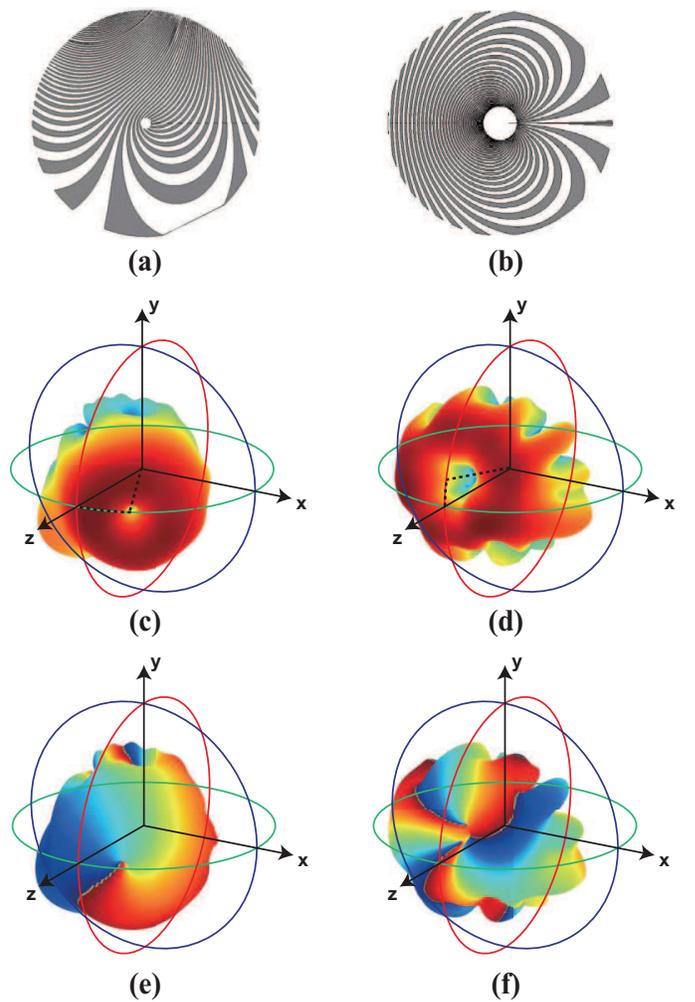}
\caption{The quasi-continuous PEC-PMC metasurfaces to produce OAM at oblique directions and their far-field radiation patterns. When $m=0.5$, $\theta_0=20^\circ$, $\phi_0=0$, (a) the top view of the PEC layer; (c) directivity and (e) phase of the RCP component. When $m=1.5$, $\theta_0=10^\circ$, $\phi_0=90^\circ$, (b) the top view of the PEC layer; (d) directivity and (f) phase of the RCP component.}
\label{q_0.5_obl}
\end{figure}

Regarding oblique reflection to be manipulated, transverse wavenumbers, $k_x$ and $k_y$ need to be introduced. They are calculated based on the desired reflected wave direction $(\theta_0,\phi_0)$:
\begin{equation}
	k_x= k_0 \cos \phi_0 \sin \theta_0, \quad k_y= k_0 \sin \phi_0 \sin \theta_0.
\end{equation}
Then, the required orientation $\theta(r,\phi)$ of metal strips is found from~\eqref{eq:theta}. By following the calculations in~\eqref{eq:k0} and~\eqref{eq:g}, the pattern of the PEC layer can be generated. Figure~\ref{q_0.5_obl} displays the reflected OAM waves at different directions. As expected, the OAM indexes are $1$ in Fig.~\ref{q_0.5_obl}(a) and $3$ in Fig.~\ref{q_0.5_obl}(b). The simulated directivity is consistent with our objective. It is worth noting that the proposed design is not limited to radio regime, as long as the responses of the PMC and PEC layers are replicated. For example, in terahertz regime, graphene-based high-impedance surfaces~\cite{yinwenyan2015} and dielectric metasurfaces~\cite{2016THZ} can be used as magnetic mirrors. In optical regime, dielectric metamaterials acting as perfect reflectors with the reflection phases of $\pi$ and zero could be applied~\cite{2015largescale}.

\section{Conclusion}
We have proposed a quasi-continuous PEC-PMC metasurface with a systematic design route to generate vortex beams at normal and oblique directions. The introduction of spatial phase to the incident plane wave is based on the concept of geometric phase. Specifically, the local phase shift depends on the orientation of the metal strips of the PEC layer. Patterning of the metal strips is accomplished with the assistance of grating vector, which offers a simple and effective way to the design of the whole metasurface. Different from existing design protocols for geometric-phase based metasurfaces, complicated optimization process of single scatterer is not needed. Furthermore, thanks to the quasi-continuous geometries, high-order diffraction from discrete scatterers are avoided.

\section*{Acknowledgment}
This work was supported in part by the Research Grants Council of Hong Kong GRF 17209918, AOARD FA2386-17-1-0010, NSFC 61271158, HKU Seed Fund 201711159228, and Hundred Talents Program of Zhejiang University under Grant No. 188020*194231701/208.

\ifCLASSOPTIONcaptionsoff
  \newpage
\fi



%

\end{document}